\documentclass[preprint2]{aastex}

\shorttitle{Empirical models for Dark Matter Halos. III.}
\shortauthors{Graham et al.}

\begin{document}

\title{Empirical Models for Dark Matter Halos. III. The Kormendy relation 
and the $\log \rho_{\rm e}$--$\log R_{\rm e}$ relation}

\author{Alister W.\ Graham\altaffilmark{1}}
\affil{Mount Stromlo and Siding Spring Observatories, Australian National 
University, Private Bag, Weston Creek PO, ACT 2611, Australia.}
\altaffiltext{1}{Graham@mso.anu.edu.au}

\author{David Merritt}
\affil{Department of Physics, Rochester Institute of Technology, 
Rochester, NY 14623, USA.}

\author{Ben Moore}
\affil{University of Zurich, Winterthurerstrasse 190, CH-8057, 
Z\"urich, Switzerland.}

\author{J\"urg Diemand}
\affil{Department of Astronomy and Astrophysics, University 
of California, 1156 High Street, Santa Cruz, CA 95064, USA.}

\author{Bal{\v s}a Terzi\'c}
\affil{Department of Physics, Northern Illinois University, 
DeKalb, IL 60115, USA.}

\begin{abstract}

\vspace{-3mm}

We have recently shown that the 3-parameter density-profile model from
Prugniel \& Simien provides a better fit to simulated, galaxy- and
cluster-sized, dark matter halos than an NFW-like model with arbitrary
inner profile slope $\gamma$ (Paper~I).  By construction, the
parameters of the Prugniel-Simien model equate to those of the
S\'ersic $R^{1/n}$ function fitted to the projected distribution.
Using the Prugniel-Simien model, we are therefore able to show that
the location of simulated ($10^{12} M_{\sun}$) galaxy-sized dark
matter halos in the $\langle \mu \rangle_{\rm e}$--$\log R_{\rm e}$
diagram coincides with that of brightest cluster galaxies, i.e., the
dark matter halos appear consistent with the Kormendy relation defined
by luminous elliptical galaxies.  These objects are also seen to
define the new, and equally important, relation $\log(\rho_{\rm e}) =
0.5 - 2.5\log(R_{\rm e})$, in which $\rho_{\rm e}$ is the internal
density at $r=R_{\rm e}$.  Simulated ($10^{14.5} M_{\sun}$)
cluster-sized dark matter halos and the gas component of real galaxy
clusters follow the relation $\log(\rho_{\rm e}) = 2.5[1 - \log(R_{\rm
e})]$.
Given the shapes of the various density profiles, we are able to
conclude that while dwarf elliptical galaxies and galaxy clusters can
have dark matter halos with effective radii of comparable size to the
effective radii of their baryonic component, luminous elliptical
galaxies can not.  For increasingly large elliptical galaxies, with
increasingly large profile shapes $n$, to be dark matter dominated at
large radii requires dark matter halos with increasingly large
effective radii compared to the effective radii of their stellar
component.

\end{abstract}

\keywords{
dark matter ---
galaxies: fundamental parameters --- 
galaxies: halos
galaxies: structure --- 
}

\section{Introduction}

Although Jaffe (1983), Hernquist (1990), and Dehnen (1993) introduced
their highly useful density-profile models to match the deprojected
form of de Vaucouleurs' (1948) $R^{1/4}$ model,
these models do not immediately yield the types of structural
quantities measured nightly by observers.
For example, observers typically refer to a galaxy's (projected)
half-light radius and surface density at this radius.  On the other
hand, modelers frequently use the Navarro, Frenk, \& White (1995,
hereafter NFW) model --- a modified version of Hernquist's model ---
and report scale radii\footnote{Due to the divergence of the mass of
the NFW model, a half-light radii cannot be defined.}  and scale
densities ($r_{-2}$ and $\rho_{-2}$) where the slope of the (internal)
density profile equals $-2$.  Or, typically, they might report on the
`concentration', related to the ratio of $\rho_{-2}$ with the average
background density of the universe, or the ratio of $r_{-2}$ with the
halo's virial radius.

Recently, alternatives to both the NFW model, and it's generalization
with arbitrary inner profile slope $\gamma$, have been shown to
provide better fits to the density profiles of simulated dark
matter halos (e.g., Merritt et al.\ 2006, hereafter Paper~I, and
references therein).
Considering fits to galaxy- and cluster-sized dark matter halos built
from hierarchical $\Lambda$CDM simulations, and fits to dark matter
halos constructed from cold spherical collapses, two density models
stand out (Paper~I).

The first is Einasto's (1965) model; see Tenjes, Haud, \& Einasto
(1994) for a more recent application.  This model has the same
functional form as S\'ersic's (1963, 1968) model but is applied to
internal (3D) density profiles rather than projected surface density
profiles.  Although we will not be using Einasto's model in this
paper, having studied it in Paper~I and Graham et al.\ (2006,
hereafter Paper~II), we note that it has recently been applied to
halos by Navarro et al.\ (2004), Diemand, Moore, \& Stadel (2004b),
and Merritt et al.\ (2005).

The second density model, by Prugniel \& Simien (1997), which {\it is}
used here, is of particular interest because it is defined using two
of the three parameters contained in S\'ersic's $R^{1/n}$ 
(light-profile) model.  Indeed, the Prugniel-Simien model was
developed to match the deprojected form of S\'ersic's model.
Specifically, the (projected) effective radius $R_{\rm e}$ and the
(projected) profile shape $n$ are common to both.  The third
parameter in Prugniel \& Simien's model, $\rho_{\rm e}$, an internal
density term at $r=R_{\rm e}$, can readily be used to obtain the
associated surface density, $\mu_{\rm e}$, at the projected radius
$R=R_{\rm e}$.  This model therefore allows one to directly compare
the structural properties of simulated dark matter halos with the
structure of real galaxies and clusters, and to do so using a better
fitting function than the generalized NFW model.
In addition, Terzi\'c \& Graham (2005) have already shown that the
Prugniel-Simien model describes the deprojected light-profiles of real
elliptical galaxies better than the Jaffe, Hernquist and Dehnen models.
%

For both real spheroidal stellar systems and simulated dark matter
halos alike, we have been able to plot their location on diagrams of
mass versus: profile shape, effective radius, effective surface
density, and effective internal (3D) density.  In the latter three
figures, the galaxy-sized dark matter halos are observed to be
consistent with the high-mass extension of ordinary elliptical
galaxies.  Moreover, the cluster-sized dark matter halos are seen to
coincide with the location of real galaxy clusters in all four
diagrams, and possibly define a trend in the $M-n$ and $M-\mu_{\rm e}$
diagrams.  This is suggestive that the same (mass-dependent) physical
processes are in operation in the real and simulated systems.

While it remains unclear as to why the simulated galaxy-sized
halos do not follow the mass versus profile shape ($M-n$) 
relation defined by real elliptical galaxies (although see Nipoti 
\& Ciotti 2006), we have been able to
use this observation to constrain the galaxy-to-halo size ratio. 
In order for (dwarf and giant) elliptical galaxies to be dark matter
dominated at large radii, we show that this ratio must increase with 
galaxy mass. 

Due to the use of the Prugniel-Simien density-profile model, we have
also been able to show, for the first time, the location of simulated
dark matter halos in the Kormendy (1977) diagram.  Within this plane
of effective surface density versus effective radius, luminous
elliptical galaxies are observed to follow the relation $\mu_{\rm e}
\sim 3\log R_{\rm e}$ (e.g., Hamabe \& Kormendy 1987; Graham 1996; La
Barbera et al.\ 2004).  Intriguingly, the galaxy-sized dark matter
halos also appear consistent with this relation.  
Furthermore, we present a new diagram, and relation, using the
internal (3D) effective density. Not only do the galaxy-sized dark
matter halos follow the same linear trend defined by luminous
elliptical galaxies, but the real galaxy clusters and cluster-sized
dark matter halos follow their own linear relation which is offset in
density by two orders of magnitude.  Such scaling laws refelct
important regulatory mechanisms that dictate the formation and growth
of galaxies and their dark matter halos, and as such these relations 
provide both inisght into these processes and useful constraints for galaxy
modelling.

In the following Section we present the mathematical form of the
Prugniel-Simien model.  In Section~3 we introduce the simulated and
real data sets which are subsequently compared with each other in
Section~\ref{results}.  Section~\ref{mass} shows trends with mass
while Section~\ref{SecComp} presents 
the $\langle\mu\rangle_{\rm e}$--$\log R_{\rm e}$ and the 
$\log\rho_{\rm e}$--$\log R_{\rm e}$ diagrams. 
A brief summary is given in Section~\ref{SecSum}.

\section{The Prugniel-Simien model: A deprojected S\'ersic model}

The density model presented in Prugniel \& Simien (1997, their
equation~B6) was designed to match the deprojected form of S\'ersic's
$R^{1/n}$ (1963) function used for describing the projected
distribution of light in elliptical galaxies.
S\'ersic's model can be written in terms of the projected intensity
profile, $I(R)$ such that
\begin{equation}
I(R)=I_{\rm e} \exp\left\{ -b_n\left[ (R/R_{\rm e}) ^{1/n} -1\right] \right\},
\label{Sersic}
\end{equation}
where $I_{\rm e}$ is the surface flux density, i.e.\ the intensity, at
the (projected) effective radius $R_{\rm e}$.  The third parameter $n$
describes the curvature, or shape, of the light profile.  The
remaining term $b_n$ is not a parameter but a function of $n$ chosen
so that $R_{\rm e}$ encloses half of the (projected) total galaxy
light.  It can be obtained by solving the expression $\Gamma
(2n)=2\times\gamma (2n,b_n)$ (e.g., Ciotti 1991, his Equation 1), 
where the quantity $\Gamma(a)$ is the gamma function and $\gamma(a,x)$ is the
incomplete gamma function given by
\begin{equation}
\gamma(a,x) = \int_0^{x}{\rm e}^{-t}t^{a-1} {\rm d}t, \hskip10pt  a>0.
\end{equation}
Although we use the exact solution for $b_n$,
a good approximation is given in Prugniel \& Simien (1997) as
\begin{equation}
b_n \approx 2n-1/3+0.009876/n, \hskip15pt n \gtrsim 0.5. 
\label{Eqbutt}
\end{equation}
A more detailed review of the S\'ersic function can be found in Graham
\& Driver (2005).

Prugniel \& Simien's (internal) density profile model can be expressed as 
\begin{equation}
  \rho(r) = \rho_{\rm e} \left({r\over R_{\rm e}}\right)^{-p}
    \exp\left\{ -b_n\left[\left( r/R_{\rm e} \right)^{1/n} -1 \right] \right\}. 
\label{EqPS97}
\end{equation}
The shape and radial scale parameters $n$ and $R_{\rm e}$ will be
recognizable from equation~\ref{Sersic}, as will the quantity $b_n$.
The third parameter, $\rho_{\rm e}$, is the internal density at the
radius $r=R_{\rm e}$.
The final quantity $p$ is not a parameter but a function of $n$ chosen
to maximize the agreement between the Prugniel-Simien model and a
deprojected S\'ersic model having the same parameters $n$ and $R_{\rm
e}$. Over the radial interval $10^{-2} \le r/R_{\rm e} \le 10^3$, a
good match is obtained when
\begin{equation}
p = 1.0 - \frac{0.6097}{n} + \frac{0.05463}{n^2}, 
          \hskip10pt 0.6\lesssim n \lesssim 10 
\end{equation}
(Lima Neto et al.\ 1999; see also Paper I, their Figure~13).

The value of $p$ is also responsible for determining the inner
logarithmic slope of the density profile (see Paper~II).  Setting
$p=0$, the Prugniel-Simien model has the same functional form as
S\'ersic's model.  When this function (with $p=0$) is applied to
density profiles, we refer to it as Einasto's (1965) model.

As noted in Lima Neto et al.\ (1999) and M\'arquez et al.\ (2001), the
associated mass profile of equation~\ref{EqPS97} is given by the
equation
\begin{equation} \label{PSmass}
M(r)  = {4\pi n {R_{\rm e}}^3} \rho_{\rm e} {\rm e}^{b_n} {{b_n}^{-(3-p)n}}
\gamma\left([3-p]n,Z\right),
\end{equation}
where $Z \equiv b_n(r/R_{\rm e})^{1/n}$.  
The total mass is obtained by replacing $\gamma(n[3-p],Z)$ 
with $\Gamma(n[3-p])$.  Expressions for the associated gravitational
potential, force, and velocity dispersion can be found in Terzi\'c \&
Graham (2005).

Equating the volume-integrated mass from equation~\ref{EqPS97} 
(i.e., the total mass from equation~\ref{PSmass}) 
with the area-integrated mass from equation~\ref{Sersic}
($=M/L\int I(R)2\pi R\, {\rm d}R$), the projected 
density, $I_{\rm e}$, at the projected radius $R=R_{\rm e}$ is given
by
\begin{equation} \label{projIe}
 I_{\rm e} = \left( {M\over L} \right)^{-1} 2 \rho_{\rm e} R_{\rm e}
  {b_n}^{n(p-1)} {\Gamma(n[3-p]) \over \Gamma(2n) }, 
\end{equation} 
The inverse mass-to-light ratio $(M/L)^{-1}$ converts the mass density into a flux
density, $I_{\rm e}$. 
New comparisons of dark matter halos (fitted with the Prugniel-Simien
model) with real galaxies (fitted with S\'ersic's model) can now 
readily be made.

\section{The Data}

\subsection{Simulated dark matter halos}

We use a sample of six cluster-sized halos (models: A09,
B09, C09, D12, E09, and F09) resolved with 5 to 25 million particles
within the virial radius, and four galaxy-sized halos (models: G00,
G01, G02, and G03) resolved with 2 to 4 million particles.
Specific details about these relaxed, dark matter halos 
formed from a hierarchical $\Lambda$CDM simulation are reported in 
Diemand, Moore, \& Stadel (2004a,b).

We have taken the profile shapes $n$ and the effective radii $R_{\rm e}$ from
the best-fitting Prugniel-Simien models applied in Paper~I. 
These quantities are equivalent to the values of $R_{\rm e}$ and $n$
obtained when fitting S\'ersic's $R^{1/n}$ model to their projected
distribution.  To obtain the halo's (projected) surface density,
$\mu_{\rm e}$, at the projected radius $R=R_{\rm e}$, we solved for
$I_{\rm e}$ in equation~\ref{projIe} to obtain 
$\mu_{\rm e} = -2.5\log(I_{\rm e})$.
%
%
Another quantity
frequently used by observers is the average (projected) surface
density within the radius $R_{\rm e}$.  It is denoted by
$\langle\mu\rangle_{\rm e}$ and given by the expression
\begin{equation}
\langle\mu\rangle_{\rm e} = \mu_{\rm e} - 2.5\log\left[
n{\rm e}^{b_n} {b_n}^{-2n}\Gamma(2n) \right]  \label{EqMuMu}
\end{equation}
(e.g.\ Graham \& Colless 1997, their Appendix A).  

Lastly, we have used the virial masses reported in Diemand et al.\ 
(2004a).  Although a standard quantity, we do note in passing
that the virial radii associated with these masses do not actually
denote the outer boundary of each halo.  For example, Prada et al.\
(2006) report on measurements out to several virial radii.  For our
sample of ten halos, the virial radii are $\sim$1.5 times larger than
the (projected) effective radii.  If we were to instead use the total
mass from equation~\ref{EqMass} below, it would be up to $\sim$2 times
larger.  The consequences of this would only influence the position of
the halo masses plotted in Figure~\ref{FigComp}.

\subsection{Real galaxies and galaxy clusters}

We have used the nearby ($z \lesssim 0.3$) 
elliptical galaxy compilation presented in Graham \&
Guzm\'an (2003).  It consists of 250 dwarf and giant elliptical galaxies
spanning a range in absolute magnitude from $-13$ to $-23$ $B$-mag.  
The bulk of these objects have had their light-profiles fitted with
S\'ersic's $R^{1/n}$ model. 
Before comparing the dark matter halo parameters with those from the stellar
distribution in real galaxies, we first had to convert the galaxy 
absolute magnitudes ($M_{\rm gal}$) into masses, and convert their
surface densities from mag arcsec$^{-2}$ to solar density per square
parsec.

We used the following simple approach to convert the $B$-band fluxes
into masses.  Each galaxy's stellar mass is simply given by
\begin{equation}
{\rm Mass} = \frac{M}{L} 10^{ 0.4\,\left(M_{\rm Sun}-M_{\rm gal}\right)}
\label{EqMass}
\end{equation}
where the $B$-band stellar (not total) mass-to-light ratio $M/L = 5.3$
(Worthey 1994, for a 12 Gyr old SSP) and the absolute magnitude of the
Sun is taken to be $M_{\rm Sun} = 5.47 B$-mag (Cox 2000).

The surface density at $R=R_{\rm e}$, denoted by $\mu_{\rm e}$, 
was transformed such that 
\begin{eqnarray}
-2.5\log(I_{\rm e} [M_{\sun} \hskip4pt {\rm pc}^{-2}])
& = &
\mu_{\rm e} [{\rm mag \hskip4pt arcsec}^{-2}] \nonumber\\
 & & \hskip-90pt
- DM - M_{\rm Sun} -2.5\log \left( \frac{M}{L} \frac{1}{f^2} \right),
\end{eqnarray}
where $DM$ is the distance modulus, equal to $25+5\log$(Distance [Mpc]), 
to each galaxy and $f=4.85\times$(Distance [Mpc]) is the number 
of parsec corresponding to 1 arcsecond at the distance of each galaxy. 
This equation subsequently reduces to
\begin{eqnarray}
-2.5\log(I_{\rm e} [M_{\sun} \hskip4pt {\rm pc}^{-2}]) & = & 
\mu_{\rm e} [{\rm mag \hskip4pt arcsec}^{-2}] \nonumber\\
 & & \hskip-90pt
- 25 - M_{\rm Sun} -2.5\log \left( \frac{M}{L} \frac{1}{4.85^2} \right). \label{mucon}
\end{eqnarray}
The internal density at $r=R_{\rm e}$, denoted by $\rho_{rm e}$, 
was derived using equation~\ref{projIe}. 

The galaxy cluster data used in this paper has come from Demarco et
al.\ (2003) who fit S\'ersic's $R^{1/n}$ model to the projected X-ray
gas distribution observed by ROSAT in two dozen clusters.  
We have used the S\'ersic scale radii\footnote{We converted their scale
radii, $a$, into effective radii using $R_{\rm e}=a(b_n)^n$.} and 
profile shapes\footnote{Note: Demarco et al.\ (2003) used $\nu = 1/n$.} 
from their Table~2,
along with the central surface densities and gas masses listed in their Table~3. 

\section{Parameter correlations}\label{results}

In this section we directly compare the structural parameters of the
($N$-body) dark matter halos, modeled in Paper~I, with the parameters
of real elliptical galaxies and real galaxy clusters.

\subsection{Trends with mass}\label{mass}

\begin{figure*}
\includegraphics[scale=0.67,angle=270]{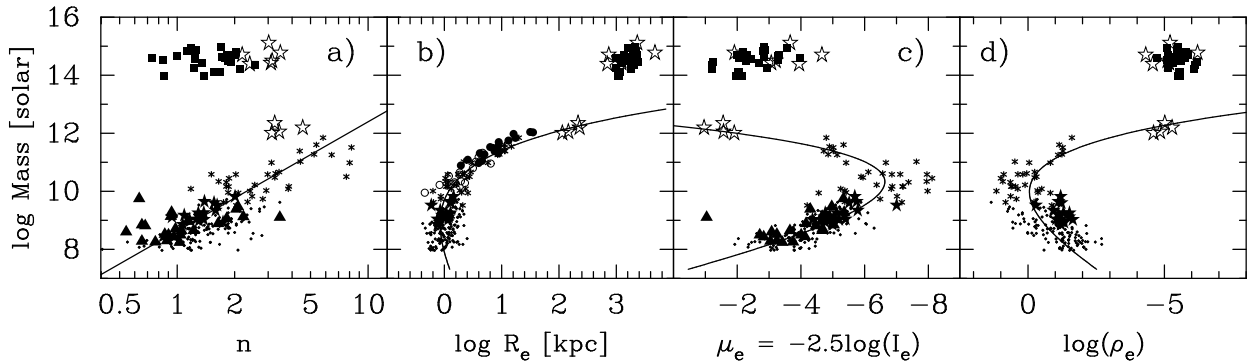}
\caption{
Mass versus a) profile shape ($n$), b) size ($R_{\rm e}$), 
c) projected surface density at $R=R_{\rm e}$, 
$\mu_{\rm e} = -2.5\log(I_{\rm e}[M_{\sun}$ \hskip2pt {\rm pc}$^{-2}$]), and 
d) spatial density at $r=R_{\rm e}$ 
($\log \rho_{\rm e}[M_{\sun}$ \hskip2pt {\rm pc}$^{-3}$]). 
The straight and curved lines are explained in the text. 
For the galaxies and galaxy clusters, the parameters have come from
the best-fitting S\'ersic $R^{1/n}$ model to the (projected) light-
and X-ray profiles, respectively.  Equivalently, the best-fitting
Prugniel-Simien model parameters to the density profiles of the 
simulated DM halos are shown.
We are are thus plotting baryonic properties for the galaxies 
alongside the dark matter properties for the halos. 
Open stars: $N$-body, dark matter halos from Paper~I; 
filles squares: galaxy clusters from Demarco et al.\ (2003); 
dots: dwarf Elliptical (dE) galaxies from Binggeli \& Jerjen (1998);
triangles: dE galaxies from Stiavelli et al.\ (2001);
filled stars: dE galaxies from Graham \& Guzm\'an (2003);
asterisk: intermediate to bright elliptical galaxies from Caon et al.\
(1993) and D'Onofrio et al.\ (1994); open and filled circles:
``power-law'' (i.e.\ S\'ersic $R^{1/n}$, see Trujillo et al.\ 2004) 
and ``core'' elliptical galaxies from Faber et al.\ (1997). 
}
\label{FigComp}
\end{figure*}

Figure~\ref{FigComp} shows the virial masses of ten $N$-body halos,
together with the stellar masses from the Graham \& Guzm\'an (2003) sample of
elliptical galaxies, and the tabulated gas masses for 24 
galaxy clusters studied in Demarco et al.\ (2003).  These masses are plotted against the
shape of the density distributions ($n$), the effective radii ($R_{\rm
e}$), the effective surface densities ($\mu_{\rm e}$), and the
internal densities, $\rho_{\rm e}$, at $r=R_{\rm e}$.  


The existence of a deviations, and a luminosity-dependent trend, 
in the shape of the light-profiles of elliptical galaxies 
has been known for two decades (e.g., 
Michard 1985; 
Schombert 1986; 
Caldwell \& Bothun 1987; 
Capaccioli 1987; 
Kormendy \& Djorgovski 1989, their Section~7.1). 
In Figure~\ref{FigComp}a we explore how this trend compares with the
structure of simulated dark matter halos.  
The galaxy-sized dark matter halos are seen to have smaller profile shapes, $n$, 
than elliptical galaxies of comparable mass.  
A further mismatch in this diagram arises from the fact that stars in
elliptical galaxies are known to have a range of distributions, i.e.\
profile shapes ($0.5 \lesssim n \lesssim \hskip4pt \sim$10; e.g.\ 
Phillipps et al.\ 1998; Caon, Capaccioli, \& D'Onofrio 1993), whereas $N$-body 
dark matter halos have shape parameters $n \sim 3\pm1$ (this discrepancy 
was previously noted by {\L}okas \& Mamon 2001 for NFW halos). 
Transforming the $B$-band absolute magnitude--$\log(n)$ relation in 
Graham \& Guzm\'an (2003, their Figure~10) into a mass relation 
using equation~\ref{EqMass} gives
\begin{equation}
\log({\rm Mass}\, [M_{\sun}]) = 8.6 + 3.8\log(n). 
\end{equation}
At odds with the slope of the $M-n$ relation for the elliptical
galaxies is the slope of the line connecting the simulated cluster-sized 
halos with the simulated galaxy-sized halos --- which has the opposite sign
(as noted by Lokas \& Mamon 2001, and also seen in Figure~4 of
Merritt et al.\ 2005).  Here, for the first time, we have included
(real) galaxy clusters in this diagram; they appear well connected with the
simulated cluster-sized halos, strengthening support for the cluster
simulations.  
It would be interesting to know where the intermediate-mass 
population ($10^{13} M_{\sun}$: galaxy groups) 
reside in this diagram. 

In Figure~\ref{FigComp}b, the scale sizes, $R_{\rm e}$, of the
galaxy-sized halos appear consistent with the (extrapolated) high-mass
end of the elliptical galaxy distribution in this diagram.  
The curved line shown in this panel has been derived from the 
$M_{\rm gal}$--$\langle\mu\rangle_{\rm e}$ relation in Graham \& Guzm\'an
(2003, the dotted curved in their Figure~12) and the relation
$L_{\rm gal} = 10^{-M_{\rm gal}/2.5} = 
2(\pi {R_{\rm e}}^2 \langle I \rangle_{\rm e})$, 
where $\langle I \rangle_{\rm e} = 10^{-\langle\mu\rangle_{\rm e}/2.5}$ 
is the average (projected) intensity within $R_{\rm e}$. 
This relation simply states that the total luminosity equals twice 
the projected luminosity inside of the effective half-light radius.

The galaxy-sized halos are also consistent with the extrapolated 
relation between mass and $\mu_{\rm e}$ for elliptical galaxies, 
shown in Figure~\ref{FigComp}c. 
As discussed in Graham \& Guzm\'an (2003), the $M_{\rm gal}$--$\mu_{\rm e}$
relation is curved, as is its mapping into the Mass--$\mu_{\rm e}$ 
plane shown here. 
The Kormendy relation, which applies to high-mass elliptical galaxies, 
has a slope of $\sim$1/4 in this diagram, and the 
%
change in slope below $\sim 10^{10}-10^{11} M_{\sun}$ is well understood
(Graham \& Guzm\'an 2003, their section~4).

The high-mass arm of the elliptical galaxy distribution in
Figure~\ref{FigComp}d reaches out to encompass the 
location of the galaxy-sized halos.  
The curved line shown there has been derived from the lines in 
Figures~\ref{FigComp}a-c together with equation~\ref{projIe}. 
%
%
Aside from the shape of the density distribution (Figure~\ref{FigComp}a), 
the `effective' parameters of galaxy-sized dark matter halos are seen
to follow the relations defined by elliptical galaxies.  That is, their
(projected)\footnote{We note that the internal radius which defines
a volume enclosing half of the mass is not equal to $R_{\rm e}$.} 
half-mass radii and the density at these radii obey the trends defined 
by the stellar mass component of elliptical galaxies. 

The structural properties of the cluster-sized halos appear largely
consistent with, or rather, for an extension to the distribution of galaxy
clusters in every panel.


\subsection{The Kormendy relation and 
the $\log \rho_{\rm e}$--$\log R_{\rm e}$ plane}\label{SecComp}

\begin{figure}
\includegraphics[scale=0.33,angle=270]{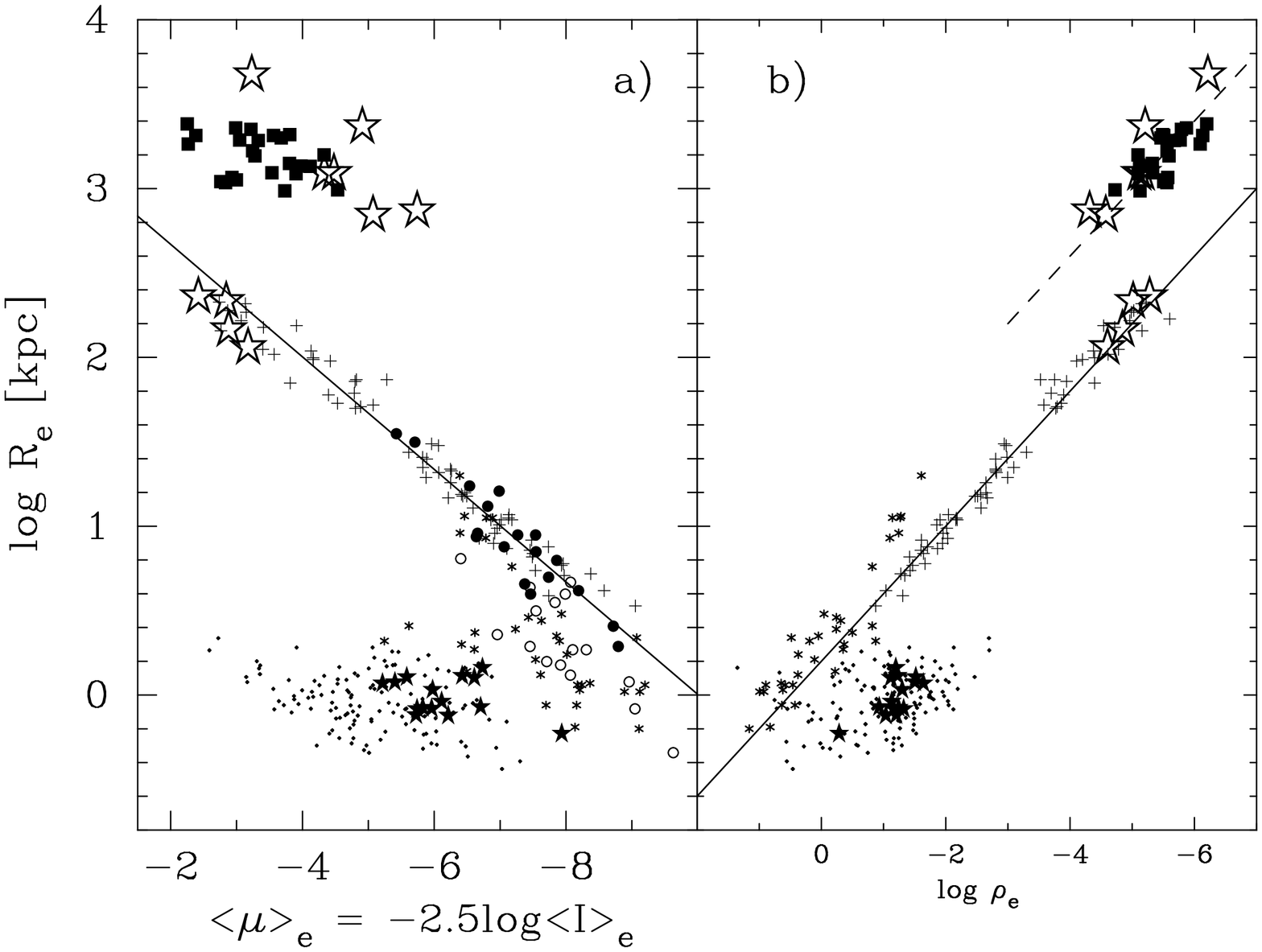}
\caption{
Effective half-mass radius versus a) the average projected density
within $R=R_{\rm e}$ (i.e.\ the mean effective surface brightness) and
b) the internal density at $r=R_{\rm e}$.  
The line in panel a) has a slope of 1/3 and roughly reproduces the
Kormendy relation that is known to hold for luminous elliptical
galaxies.  Lines of constant mass have a slope of 1/5 in panel a).
The new relations in panel b) are given by 
$\log R_{\rm e} = 0.2 - 0.4 \log \rho_{\rm e}$ for the galaxy-sized
objects and 
$\log R_{\rm e} = 1.0 - 0.4 \log \rho_{\rm e}$ for the cluster-sized objects. 
The symbols have the same meaning as in Figure~\ref{FigComp}, with
additional filled squares denoting the brightest cluster galaxies from
Graham et al.\ (1996).
}
\label{FigKorm}
\end{figure}

Figure~\ref{FigKorm}a shows the effective radius, $R_{\rm e}$, versus the
average (projected) surface density inside of $R_{\rm e}$,
$\langle\mu\rangle_{\rm e}$.  We have been able to augment this
diagram with the brightest cluster galaxy (BCG) sample from Graham et
al.\ (1996, their Table~1) that were fitted with S\'ersic's $R^{1/n}$
model.  This required 
converting their $R$-band surface brightness data to the $B$-band
using the average color $B-R = 1.57$ (Fukugita, Shimasaku, \& Ichikawa
1995),
%
%
applying equation~\ref{mucon} to obtain the surface density at $R_{\rm e}$ 
in units of solar masses per square parsec, and then 
deriving $\langle\mu\rangle_{\rm e}$ from $\mu_{\rm e}$ using 
equation~\ref{EqMuMu}.  
(No reliable stellar masses exist for these galaxies.)
We note that the effective radii of the BCGs with large S\'ersic
indices are, in some instances, greater than the observed radial
extent of the BCG.  As such, these scale radii are reflective of the
S\'ersic $R^{1/n}$ model which matches the observed portion of the
galaxy.  Subject to how the outer profiles truncate, these radii may
or may not represent the actual half-light radii.
The solid line in Figure~\ref{FigKorm}a has a slope of 1/3, 
typical of the Kormendy relation for luminous elliptical galaxies.
The departure of the lower-luminosity elliptical galaxies from this
relation is explained in Graham \& Guzm\'an (2003, their Section~4, 
see also Capaccioli \& Caon 1991 and La Barbera et al.\ 2002). 

The apparent agreement between the galaxy-sized dark matter halos and
the BCGs implies that, within their respective effective radii, the average
projected mass density in stars (in the case of the BCGs) and in dark
matter (in the case of the halos) is equal.  It will be of interest to
see if less massive (dwarf-galaxy-sized) halos follow the Kormendy relation to higher
densities (to the right in this figure) 
or depart from this relation\footnote{We remind readers that
surface density is given by $-2.5\log$(column density per unit area),
and thus more negative numbers reflect an {\it increased} density.}.

Figure~\ref{FigKorm}b shows $\rho_{\rm e}$, the internal 
density at $r=R_{\rm e}$, versus $R_{\rm e}$.  The obvious
relation for the luminous elliptical galaxies and the galaxy-sized 
dark matter halos is such that
\begin{equation}
\log(\rho_{\rm e}) = 0.5 - 2.5\log(R_{\rm e}), 
                     \hskip10pt \log(R_{\rm e}) \gtrsim 0.5, 
\end{equation}
where $R_{\rm e}$ is in kpc and $\rho_{\rm e}$ is in units of solar
masses per cubic parsec.  
It is noted that this only describes the pan-handle of a more complex 
distribution seen in this figure, but is, we feek, nonetheless of interest. 

\begin{figure}
\includegraphics[scale=0.63,angle=270]{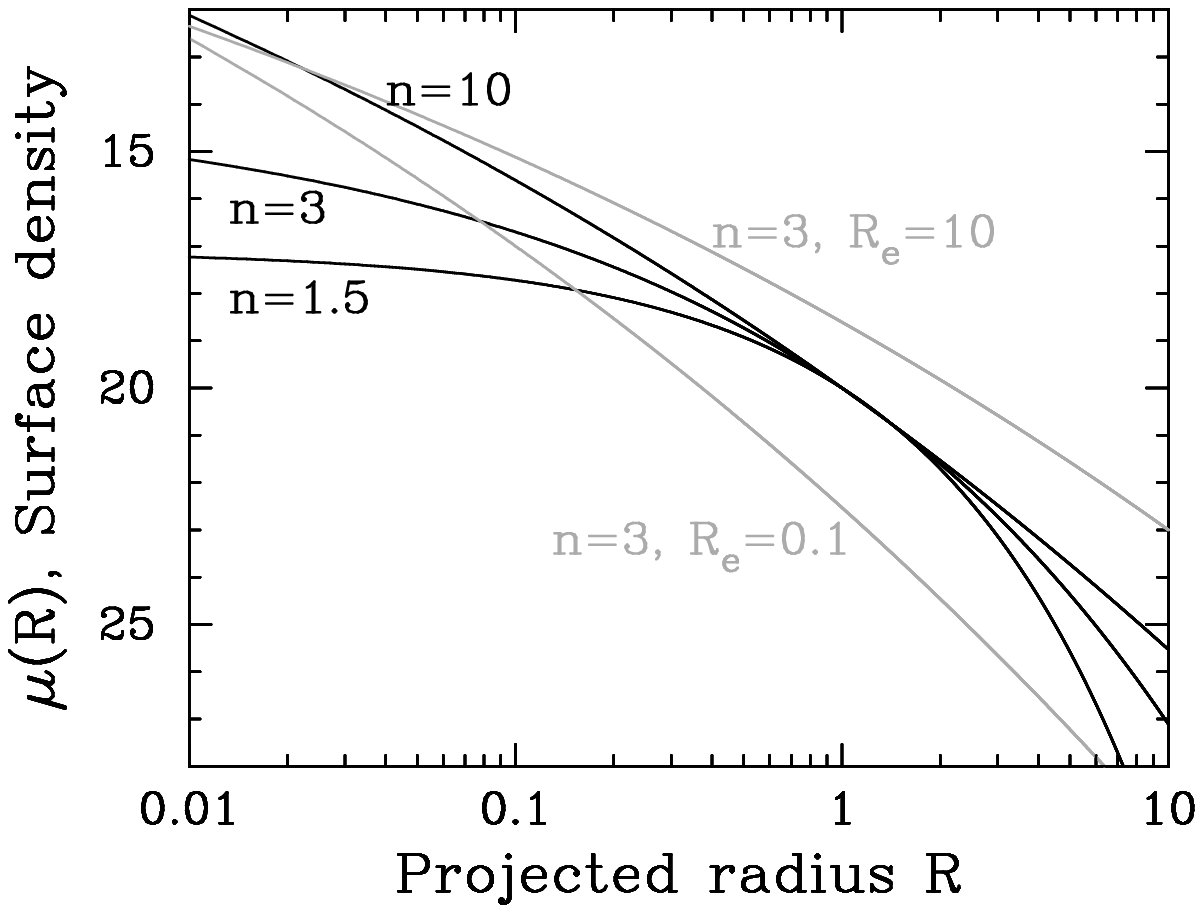}
\caption{The dark curves show three S\'ersic profiles with different shape 
parameters $n$ but similar effective radii $R_{\rm e}=1$
and effective surface densities $\mu_{\rm e}=20$. 
One can see how an $n=1.5$ galaxy cluster can have the
same $R_{\rm e}$ as an $n=3$ dark matter halo but yet be
dark matter dominated at all radii.  A luminous $n=10$
elliptical galaxy cannot be dark matter dominated at large (or any)
radii if it has an $n=3$ dark matter halo with equal stellar-to-dark
matter effective radii. 
The faint curves show two $n=3$ S\'ersic profiles, one which has 
$R_{\rm e}=0.1$ and $\mu_{\rm e}=17$, the other has 
$R_{\rm e}=10$ and $\mu_{\rm e}=23$.  The shifts in $\mu_{\rm e}$
are dictated by the change in $R_{\rm e}$ and the Kormendy relation
with a slope of 1/3. 
}
\label{FigSer}
\end{figure}

Of course, the above relation would only imply an equal
stellar-to-dark matter density ratio at $R_{\rm e}$ in galaxies {\it
if} the stellar and dark matter components had the same value of
$R_{\rm e}$.  This situation, however, can not exist if large
elliptical galaxies (with large values of $n$) are to have dark matter
halos (with $n\sim$3) 
that dominate the mass at large radii, as suggested by, for 
example, the analysis of X-ray halo gas by Humphrey et al.\ (2006)
and Sansom et al.\ (2006). 
This predicament is illustrated in Figure~\ref{FigSer}.  
%
%
The obvious, but previously unstated, answer is that {\it dark matter halos
in large elliptical galaxies must have {\it larger} effective radii
than the stellar distribution's effective radii.}
On the other hand, given that the profile shapes of dwarf elliptical
galaxies typically have S\'ersic indices less than 2
(Figure~\ref{FigComp}a), {\it a dwarf galaxy can have a dark matter halo
(with $n\sim$ 3) that has the same effective radius as the stellar
component}, and still be dark-matter dominated at all radii.  The above
situation can be visualized in Figure~\ref{FigSer}.

The simulated dark matter cluster-sized halos and the gas component of
real galaxy clusters also appear to have similar structural
properties in Figure~\ref{FigKorm}.  Specifically, in Figure~\ref{FigKorm}b,
the two populations reside in a similar part of the diagram and possibly
define their own relation offset by a factor of $\sim$6 in $R_{\rm
e}$, or $\sim$100 in density from that defined by the galaxies and
galaxy-sized halos.  Their distribution is traced by the relation
\begin{equation} 
\log(\rho_{\rm e}) = 2.5[1 + \log(R_{\rm e})],
                     \hskip10pt \log(R_{\rm e}) \gtrsim 1.5. 
\end{equation}
Similarly with dwarf elliptical 
galaxies, the low-$n$ profile shapes (see Figure~\ref{FigComp}a) 
of galaxy clusters means that
their stellar-to-dark matter effective radii can be comparable 
with the dark matter still dominating at all radii, expect, 
of course, inside a cluster's centrally-located BCG. 

The tabulated dynamical masses in
Demarco et al.\ (2003) are roughly five times greater than the
tabulated gas masses that we have used.  Similarly, using their
equation~5 to obtain the (internal) dark matter-to-gas density ratio
at $r=R_{\rm e}$, one also obtains an average value around five.
We do not, however, know the effective radii of these cluster's 
dark matter components and so we can not show where they 
reside in Figure~\ref{FigKorm}.

\section{Summary} \label{SecSum}

Simulated galaxy-sized dark matter halos appear largely consistent
with the location of brightest cluster galaxies in the 
$\langle\mu\rangle_{\rm e}$--$\log R_{\rm e}$ plane. Indeed, the halos appear 
congruent with the Kormendy relation.  Interestingly, the galaxy-sized 
halos also appear to follow the same relation as luminous elliptical 
galaxies in the $\log \rho_{\rm e}$--$\log R_{\rm e}$ plane, defining 
a new relation $\log (\rho_{\rm e}[M_{\sun} \hskip4pt {\rm pc}^{-3}]) = 0.5 - 
2.5\log (R_{\rm e}[{\rm kpc}])$.
Using this information, coupled with knowledge of the stellar and 
density profile shapes, we are able to make statements about the
relative effective radii of stellar-to-dark matter distributions
in galaxies and clusters. Specifically, while large elliptical 
galaxies require a small ratio of their stellar-to-dark matter effective
radii, dwarf 
galaxies and galaxy clusters could have a size ratio of unity 
or larger and still be dark matter dominated. 
The galaxy clusters and simulated cluster-sized dark matter halos
appear to define a new relation given by 
$\log (\rho_{\rm e}[M_{\sun} \hskip4pt {\rm pc}^{-3}]) = 2.5 - 
2.5\log (R_{\rm e}[{\rm kpc}])$.

\acknowledgments  

We kindly thank Gary Mamon for his detailed comments on this manuscript. 
A.G.\ acknowledges support from NASA grant HST-AR-09927.01-A from the
Space Telescope Science Institute, and the Australian Research Council
through Discovery Project Grant DP0451426.
D.M.\ was supported by grants AST 02-06031, AST 04-20920, and AST
04-37519 from the National Science Foundation, and grant NNG04GJ48G
from NASA.
%
%
J.D.\ is grateful for financial support from the Swiss National
Science Foundation.
B.T.\ acknowledges support from Department of Energy grant G1A62056.

\end{document}